\documentclass[twocolumn,10pt,a4paper]{article}

\usepackage[utf8]{inputenc}
\usepackage[T1]{fontenc}
\usepackage{amsmath}
\usepackage[varg]{txfonts}
\usepackage[a4paper,left=2.1cm,right=2.1cm,top=2.4cm,bottom=2.6cm,columnsep=6mm,headheight=14pt]{geometry}
\usepackage{graphicx}
\usepackage{natbib}
\bibpunct{(}{)}{;}{a}{}{,}
\usepackage{booktabs}
\usepackage{microtype}
\usepackage[font=small,labelfont=bf,labelsep=period,tableposition=top]{caption}
\usepackage{titlesec}
\titleformat*{\section}{\large\bfseries}
\titlespacing*{\section}{0pt}{2.2ex plus .6ex minus .2ex}{1.2ex plus .2ex}
\usepackage{fancyhdr}
\usepackage[hidelinks]{hyperref}
\usepackage{orcidlink}

\graphicspath{{./}}

\newcommand{\snrpink}{\ensuremath{\mathrm{S/N_{pink}}}}
\newcommand{\tabnotes}[1]{\par\vspace{3pt}\begin{minipage}{\linewidth}\footnotesize\textbf{Notes.} #1\end{minipage}}

\title{Habitable-zone Earths at the detection frontier: Measured completeness
and false-alarm rate of a transit pipeline for the PLATO M-dwarf sample}
\author{Yohann Tschudi\,\orcidlink{0009-0002-7797-1502}\\[3pt]
\normalsize Independent Researcher, 69480 Anse, France\\
\normalsize\texttt{yohann.tschudi@gmail.com}}
\date{}

\begin{document}
\twocolumn[
\begin{@twocolumnfalse}
\maketitle
\begin{abstract}
The PLATO mission will observe its first long-pointing field, LOPS2 (Long-duration
Observation Phase, South field), for at least two years, bringing the habitable
zones (HZ) of its M-dwarf sample within reach of a transit search.
The completeness and false-alarm rate of a pipeline validated on TESS
M dwarfs are measured on simulated PLATO photometry and converted into a
candidate-level forecast for the LOPS2 M dwarfs.
Simulated PLATO light curves (717~d baseline; three stellar cells of the
faint half of the sample) were processed end to end in two arms,
without and with a pre-launch model of the instrument systematics: 930 in the
first (465 realisations, 525 injected planets, 465 noise-only twins) and
790 in the second. Injected planets measure completeness against the expected
signal-to-noise ratio (S/N), separately for the long-period regime of the HZ;
noise-only twins measure the false-alarm rate against the floor on the
red-noise-aware statistic \snrpink.
Completeness reaches 50\% at ${\rm S/N}=7.7\pm0.15$ over all periods and
$8.6\pm0.3$ in the long-period regime. The false-alarm rate falls from
$79\pm2$\% to 0.6\% as the \snrpink\ floor rises from 6 to 8, and
$6\pm1$\% at the adopted floor 7.5 (completeness $54\pm2$\%). In the field, the
50 known transiting objects of interest (19 confirmed planets, 31 candidates)
would all be recovered; a transiting HZ planet of
$1.0\,(1.5)\,R_\oplus$ would be recovered as a candidate with $\sim$40\%
(65\%) probability, giving about 4 HZ Earths and 7 super-Earths over
16\,929 field M dwarfs. The systematics arm shifts the 50\% point by
$+0.54^{+0.27}_{-0.22}$ in S/N, a penalty localised on the transition, leaving the floor robust and
negligible on the long-period injections that feed the HZ.
Habitable-zone Earths around the PLATO M dwarfs sit at the measured detection
frontier, where completeness and reliability, both measurable before
launch, decide the yield.
\end{abstract}
\begin{quote}\small\noindent\textbf{Keywords.} planets and satellites: detection;
planets and satellites: terrestrial planets; techniques: photometric; stars: low-mass
\end{quote}
\vspace{1ex}
\end{@twocolumnfalse}
]

\section{Introduction}\label{sec:intro}

The PLAnetary Transits and Oscillations of stars mission
\citep[PLATO;][]{Rauer2025} will observe its first long-pointing field, LOPS2
\citep[Long-duration Observation Phase, South field;][]{Nascimbeni2025},
continuously for at least two years. Its cool-dwarf
sample \citep[P4;][]{Prisinzano2026} extends the transit search to M dwarfs,
for which the habitable zone (HZ) corresponds to orbital periods of 19--112~d
for M4 to M2 hosts; the 717~d baseline therefore accumulates 6--37 HZ transits
per target.

Yield estimates for PLATO exist along three lines: injection and recovery of
Earth-like planets around Sun-like stars \citep{Heller2022}, mission-wide
statistical yields for the FGK samples \citep{Matuszewski2023}, and
demographic survey simulations that reach the M dwarfs \citep{Schlecker2024}.
The present work complements these forecasts with the two quantities that
injection and recovery through a search-and-screening pipeline can provide: the
fraction of injected planets the pipeline returns as candidates from realistic
light curves, and the rate at which the same pipeline returns a candidate from
light curves that contain no planet. This paper reports both for the
cool-dwarf sample; the quoted yields are candidate-level, before the
astrophysical false-positive vetting that requires imaging or spectroscopy.

The pipeline was developed and validated on M dwarfs observed by the
Transiting Exoplanet Survey Satellite \citep[TESS;][]{Tschudi2026a}. Applied
uniformly to the 461 M-dwarf hosts of TESS objects of interest (TOIs) listed
by the Exoplanet Follow-up Observing Program (ExoFOP), it re-detected 165 of
the 193 confirmed planets in its period range (85.5\%) and misclassified none
of them as a false positive \citep{Tschudi2026b}.

\section{Simulated campaign and pipeline}\label{sec:campaign}

\begin{table*}[!t]
\centering
\caption{Campaign cells, shared by both arms: content and purpose.}
\label{tab:cells}
\small
\setlength{\tabcolsep}{5pt}
\begin{tabular}{@{}llrlp{5.6cm}@{}}
\toprule
Cell & Host & Noise$^{(a)}$ & Content (injections $+$ twins) & Measures \\
\midrule
M4 bright & M4, $V=14.5$, 12 cam & 1308 & 145 single planets $+$ 145 & completeness transition at the low-noise anchor; HZ band 19--52~d \\
M2        & M2, $V=15.0$, 12 cam & 2076 & 145 single planets $+$ 145 & earlier-type host: HZ band 43--112~d, the long-period extreme (6--17 events) \\
M4 faint  & M4, $V=16.0$, 6 cam  & 6200 & 145 single planets $+$ 145 & faint six-camera limit; anchors the noise model at the hard end \\
L\,98-59 analogue & M4, $V=14.5$, 12 cam & 1308 & $10\times3$ planets $+$ 10 & iterative multiplanet search; outer pair at a 2.02:1 ratio (harmonic confusion) \\
TOI-700 analogue  & M2, $V=15.0$, 12 cam & 2076 & $10\times4$ planets $+$ 10 & long-period multiplicity up to the HZ planet of the real system \\
Two-planet        & M4, $V=14.5$, 12 cam & 1308 & $10\times2$ planets $+$ 10 & near 3:2 pair; mutual masking of close signals \\
\bottomrule
\end{tabular}
\tabnotes{$^{(a)}$~ppm per 300~s bin. Single-planet injection periods span
0.5--100~d. The multiplanet periods are 2.25/3.69/7.45~d (L\,98-59 analogue),
10.0/16.1/27.8/37.4~d (TOI-700 analogue), and 2.44/3.79~d (two-planet).}
\end{table*}

Light curves were generated with the PLATO Solar-like Light-curve Simulator
\citep[PSLS;][]{Samadi2019} at the 25~s cadence of the P4 data products
\citep{Rauer2025} and binned to 300~s for the search. The binning divides the
search cost by an order of magnitude ($2\times10^{5}$ instead of
$2.5\times10^{6}$ points per 717~d light curve) at a transit S/N cost below
1\%, the bins remaining much shorter than any P4 transit duration. The native
25~s cadence remains available for the per-candidate vetting.

The simulated signal contains photon and detector noise, the nominal gap
structure, and a rotational modulation from three spots of finite lifetime; the
second arm adds the residual systematics described below. Granulation and the
stochastic activity term were disabled, being calibrated on solar-like and
evolved stars, and solar-like oscillations were disabled because the modes of
main-sequence M dwarfs lie far above the 1667~$\mu$Hz Nyquist frequency of the
300~s bins. Flares were not simulated, the most consequential omission for an
M-dwarf sample (Sect.~\ref{sec:limitations}).

Three stellar cells anchor the campaign: an M4 dwarf at $V=14.5$ observed with
12 cameras (1308~ppm per 300~s bin), an M2 at $V=15.0$ with 12 cameras
(2076~ppm), and an M4 at $V=16.0$ with 6 cameras (6200~ppm). These cells span
the faint half of the sample. The median magnitude of the field M dwarfs is
$V=15.5$; at these magnitudes the expected S/N of HZ planets of
1--1.5~$R_\oplus$ falls on the completeness transition, where recovery drops
from likely to unlikely, while brighter stars sit on the saturated plateau.
The injections therefore concentrate on the transition, as in the TESS survey
at the stellar noise frontier \citep{Tschudi2026a}.

Throughout, and for the field stars of Sect.~\ref{sec:forecast}, the expected
S/N is $\delta_{\rm win}\sqrt{n_{\rm in}}/\sigma$, with $\sigma$ the white noise
per 300~s point of the cell, $n_{\rm in}=(717\,{\rm d}/P)(T_{14}/300\,{\rm s})$
the number of in-transit bins, and $\delta_{\rm win}$ the limb-darkened window
depth, which exceeds the geometric $(R_p/R_\star)^2$ by a measured factor
$1.20$; using the geometric depth would displace the axis by 20\%. The 465 planet-hosting
realisations carry 525
injected planets with expected S/N concentrated on that transition (3.5--13);
three multiplanet configurations complete the campaign. Each realisation has a
noise-only twin, the same star and noise class without a planet; the 465 twins
carry the entire false-alarm measurement. Table~\ref{tab:cells} summarises the
cells, their content, and the quantity each one measures; the systematics arm
reuses the same cells with fewer trials.

The campaign comprises two arms, matched in cell design, S/N sampling, and
chain (435 and 371 single-planet trials, respectively). The arm reported here (arm A)
contains no residual instrument systematics and measures the intrinsic
performance of the pipeline against stellar and photon noise; the systematics
arm (arm B) injects the pre-launch model of the post-correction residuals
(Sect.~\ref{sec:limitations}) and measures their cost. The two arms share
every input except the systematics, so their difference is the instrumental
cost, and the pair reads the pipeline's intrinsic performance and the
instrumental penalty separately.

The pipeline is that of the TESS survey \citep{Tschudi2026a,Tschudi2026b}. The
chain is: Gaussian-process detrending; an iterative transit-least-squares (TLS)
search \citep{Hippke2019}, whose internal acceptance requires a signal
detection efficiency ${\rm SDE}\geq7$; harmonic-alias resolution; and an
event-time-coherence test that rejects signals whose per-transit timing is
incoherent (window-function aliases of stellar variability). Three settings
differ from the published TESS configuration.

First, the search noise is estimated by TLS, the PSLS uncertainties being
photon-only. Second, a veto rejects candidates within $0.5$\% of the quarterly
rotation comb ($90.004$~d and low harmonics), the PLATO analogue of the
$13.7$~d entry used for TESS: the rotations imprint structures repeating at the
quarter period, an instrumental clock rather than a transit. It shapes the
false-alarm rate below and costs three injections, none recovered, one at
${\rm S/N}=11.8$ and $P=29.90$~d that the chain would otherwise have detected;
the comb crosses the HZ band of the M2 cell, leaving three narrow blind periods.
Third, a power gate on the event-time test lets it abstain rather than reject
when the median per-event depth significance falls below $0.85\sigma$, a
threshold inside an interval left empty by the calibration set (noise twins
reach $0.39$, false rejections $0.61$, confirmed planets begin at $1.13$, and
the certified window-function false positive at $1.18$ stays rejected). The gate
changes 29 of 435 verdicts, all at $P\leq10.6$~d and none in the HZ band, where
recovery is 37 of 75 with and without it, so the HZ forecast is independent
of it. Every scored candidate carries an S/N computed from the white and
red noise contributions at the transit timescale, noted \snrpink\ after its
pink-/red-noise formulation \citep{Pont2006}; the operating points reported
below apply a floor on this statistic.

\section{Completeness and false-alarm rate}\label{sec:results}

\begin{figure}
\centering
\includegraphics[width=\hsize]{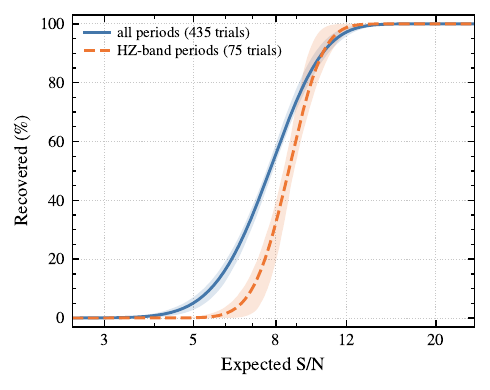}
\caption{End-to-end completeness versus expected S/N (no-systematics arm).
Blue: all periods (435 planets; the fits use the single-planet injections
only, the 90 planets of the multiplanet configurations being fixed-period
architecture tests, see Table~\ref{tab:cells}), 50\% at
${\rm S/N}=7.72^{+0.15}_{-0.14}$. Orange: long-period (HZ-band) regime (periods
inside the conservative HZ band of the host cell, 75 planets), 50\% at
${\rm S/N}=8.61^{+0.33}_{-0.28}$. Bands are bootstrap 1$\sigma$ intervals of
the fitted $\Gamma$ cumulative distribution function (CDF).}
\label{fig:completeness}
\end{figure}

Figure~\ref{fig:completeness} presents the completeness measurement. Over all
periods the recovery fraction reaches 50\% at an expected
${\rm S/N}=7.72^{+0.15}_{-0.14}$ and exceeds 90\% above ${\rm S/N}\approx9$
(94\%, 165 of 176, at ${\rm S/N}\geq9$; 96\%, 124 of 129, at $\geq10$). The
50\% point falls near the canonical ${\rm S/N}\approx7$ detection threshold of
the transit literature, here measured end to end on correlated noise.

Restricted to planets whose periods fall inside the conservative HZ band of
their host cell \citep{Kopparapu2013}, the curve shifts by $0.9^{+0.8}_{-0.5}$
in S/N (50\% at $8.61^{+0.33}_{-0.28}$; $p=0.012$ from a permutation test
reassigning the HZ labels at random and refitting the shift). These
injections still carry many transits (7--37, median 18); the shift is a
property of long-period signals rather than of the HZ, as non-HZ planets with
$P\geq19$~d show the same displacement. It arises because HZ periods
give longer transits, which the Gaussian-process detrending and correlated
noise erode more strongly. The forecasts of Sect.~\ref{sec:forecast} use this
long-period curve.

\begin{figure}
\centering
\includegraphics[width=\hsize]{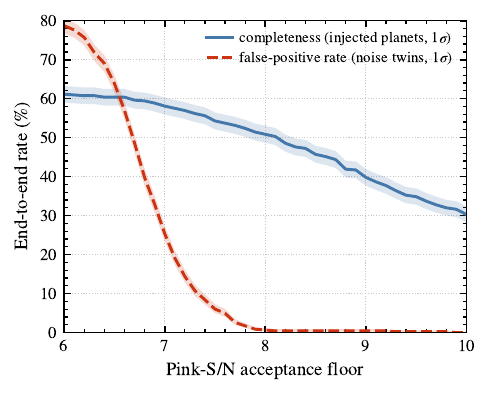}
\caption{End-to-end completeness (blue) and noise-twin false-alarm rate (red,
dashed) versus the \snrpink\ acceptance floor, with Wilson 1$\sigma$ bands
(no-systematics arm). The false-alarm rate falls from $79\pm2$\% at a floor of
6 to 0.6\% at 8; completeness decreases from $61\pm2$\% to $51\pm2$\% over the
same range.}
\label{fig:floor}
\end{figure}

The false-alarm rate is measured on the noise-only twins
(Fig.~\ref{fig:floor}). It falls from $79\pm2$\% to 0.6\% (3 of 465) as the
\snrpink\ floor rises from 6 to 8, while completeness decreases from
$61\pm2$\% to $51\pm2$\%: noise is confined below
${\snrpink}\approx7.5$--8 whereas injected planets extend above. A floor at
7.5 yields a $6.0\pm1.1$\% false-alarm rate at $54\pm2$\% completeness.
Counting only candidates with a coherent-clock verdict from the event-time
test, the rate is 1.9\% (9 of 465) for the frozen chain and 1.5\% (7 of 465)
with the power gate. On correlated noise, part of the
apparent completeness at low floors consists of false alarms; the two
quantities are therefore reported as a pair at each operating point.

\section{Application to the LOPS2 field}\label{sec:forecast}

The forecast separates measured from modelled quantities. The completeness
curves are measured on the campaign cells. Each real star then requires only
the placement of its (real or hypothetical) planets on the S/N axis, using the
same S/N convention as the campaign. The per-point noise is interpolated
between the three measured cells for the faint half of the sample and
extrapolated brightward for the rest (interpolation residuals below 3\%); stellar parameters derive from Gaia colours calibrated on the 22 field hosts with ExoFOP
parameters; and the camera count of each star follows from the four-group
field geometry, validated against the published field (union area reproduced
to 0.05\%; 13/11/33/43\% of the field covered by 24/18/12/6 cameras).

\begin{figure}
\centering
\includegraphics[width=\hsize]{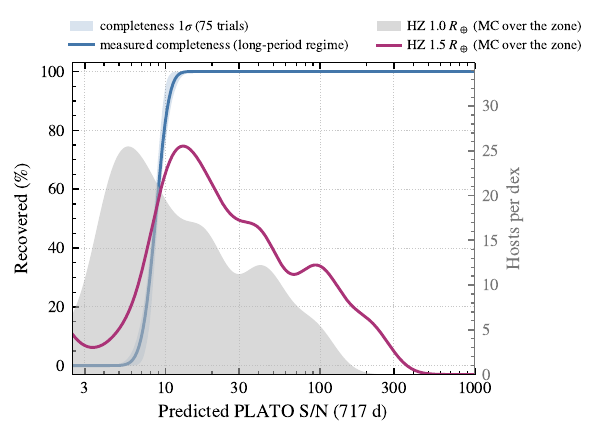}
\caption{Measured long-period completeness (blue, with its bootstrap 1$\sigma$
band; left axis) applied to the real field population (no-systematics arm).
Right axis, hosts per dex: predicted PLATO S/N distribution of habitable-zone
planets of $1.0\,R_\oplus$ (grey fill) and $1.5\,R_\oplus$ (purple), Monte
Carlo sampled over the conservative HZ of each of the 35 LOPS2 M-dwarf TOI
hosts (of the 36 hosts, the 35 with a catalogued stellar radius),
with per-star camera counts. The integrated detection probabilities quoted in
the text weight each population by the completeness curve.}
\label{fig:hz}
\end{figure}

Three applications follow. First, the objects already known: the 50 transiting
TESS objects of interest around the 36 M-dwarf hosts inside the field
footprint, 19 of them confirmed planets and 31 still candidates in the
2026-06-29 ExoFOP snapshot, all have predicted recovery probabilities at unity
(median predicted ${\rm S/N}=96$); once PLATO data exist, this population
provides an on-sky validation sample with known ephemerides. Second, the HZ benchmarks
(Fig.~\ref{fig:hz}): a transiting $1.0\,R_\oplus$ planet receiving
conservative-HZ irradiation around the same real stars would be recovered as a
candidate with an integrated probability of $41\pm1$\% (statistical only; see
below), rising to $65\pm1$\% at $1.5\,R_\oplus$. The bright cohort of the
field (L\,98-59, TOI-270, and TOI-700 are members) is recovered at either
radius; the faintest hosts are out of reach at any floor. This estimate carries
one extrapolation: the completeness curve is measured up to $P=100$~d, whereas
about a third of the real HZ hosts reach $100$--$160$~d, only 4--6 transits over
the baseline. That is the genuine few-transit regime the campaign did not
sample, where the long-duration argument above no longer holds, so the forecast
is optimistic for that third. Third, the field-population scale: over the 16\,929
M dwarfs of the PLATO input catalogue \citep{Montalto2021} inside the
footprint, with conservative-HZ occurrence rates \citep{Dressing2015}, about
19 transiting HZ Earths (1--1.5\,$R_\oplus$; 10--38 over the occurrence-rate
interval) are statistically present, of which the pipeline would recover
$4.0^{+4.2}_{-1.8}$ as candidates (68\% interval propagated from the occurrence
rates), plus $6.6^{+5.4}_{-2.8}$ HZ super-Earths (1.5--2\,$R_\oplus$). That
denominator is a population count inside the footprint, not the list of stars
P4 will observe, and the yield scales linearly with the observed fraction. Most
of the reservoir lies at ${\rm S/N}=2$--6 around stars too faint for any method.
These are candidate counts and carry the reliability of
Sect.~\ref{sec:results}: $6.0\pm1.1$\% false alarms at the adopted floor,
$1.5$--$1.9$\% once a coherent-clock verdict is required. For comparison, the
consortium performance assessment \citep{Cabrera2026} quotes ten planets below
$2\,R_\oplus$ in the P4 habitable zone at an intermediate occurrence rate,
from sensitivities \citep{Eschen2024} stated there for the P1, P5 and Prime
samples, against the $10.6$ recovered here on different stellar and occurrence
assumptions.

\section{Limitations}\label{sec:limitations}

\begin{figure}
\centering
\includegraphics[width=\hsize]{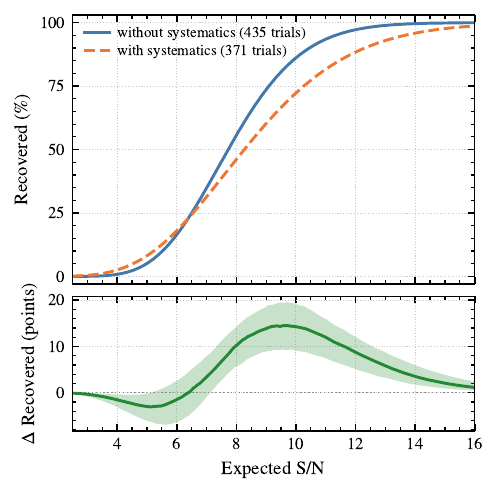}
\caption{The systematics penalty. \emph{Top}: end-to-end completeness without
(blue) and with (orange, dashed) the modelled instrument systematics, both as
$\Gamma$-CDF fits. \emph{Bottom}: their difference in recovered percentage
points, with its bootstrap 1$\sigma$ interval.}
\label{fig:bracket}
\end{figure}

The completeness, reliability, and yields of Sects.~\ref{sec:results} and
\ref{sec:forecast} are the no-instrument-systematics bound: PSLS stellar and
random noise with a perfectly corrected instrument. The systematics arm, matched
to it but with the pre-launch model of the post-correction residuals injected,
measures the cost of that assumption (Fig.~\ref{fig:bracket}): the 50\% completeness point shifts
by $+0.54^{+0.27}_{-0.22}$ in S/N (7.72 to 8.26), a loss localised on the transition (peaking at
about 14 points near ${\rm S/N}=10$) and vanishing on the plateau.

The penalty is not uniform. The residual template is transplanted at fixed
amplitude from the single configuration for which official tables exist: the
tables stop at magnitude 13, short of the campaign's cells at $V=14.5$--16,
where the random-noise term dominates the error budget. Its effect therefore
scales with the amplitude-to-noise ratio of each star: $+1.5$ at
$2.9\sigma$ on the quietest cell (1308~ppm), $+0.6$ at $1.2\sigma$ on the
intermediate one, and consistent with zero ($-0.5$ at $1.1\sigma$) on the
noisiest (6200~ppm); the quietest-to-noisiest contrast is itself significant at
$3.0\sigma$. This gradient is a
property of the injection, not a prediction of the true per-star penalty, which
would need an amplitude-versus-magnitude law the tables do not provide; on
faint stars the real residuals worsen with charge-transfer inefficiency and
would likely reverse the near-zero point.

The acceptance floor is robust. At 7.5 the false-alarm rate is $3.5\pm0.9$\% (below
the no-systematics $6\%$, as the residuals push noise candidates below the
floor) for $46\pm2$\% completeness. The HZ-Earth yield is unchanged within the
reproducibility of the estimate: the long-period injections that feed it lie in
the low-penalty regime, and the arm B completeness curve there is statistically
identical to that of arm A. Two omissions push the penalty in opposite directions: the pipeline has
no dedicated systematics-correction stage, which overstates the cost, while the
residuals are treated as camera-independent, which understates it (common-mode
terms correlate within a group). The $+0.54$ is therefore an estimate at the
modelled residual level, not a one-sided bound.

Four fidelity limits of PSLS \citep{Samadi2019} apply.
\emph{(i)}~Its stellar models are solar-like; at the campaign's faint
magnitudes the budget is photon- and instrument-dominated, and stellar
activity is a fixed spot parameterisation. Flares were not simulated
(Sect.~\ref{sec:campaign}), which makes completeness and false-alarm rate alike
optimistic by an amount no control arm bounds here.
\citet{Yaptangco2025}, applying the same family of chain to real TESS
photometry of M dwarfs, find the 50\% threshold moving from ${\rm S/N}=7\pm1$
on inactive hosts to $14\pm1.5$ on active ones: the present completeness
describes the quiet end of the population.
\emph{(ii)}~Its activity is one frozen spot configuration, the same three
spots of 75, 90, and 60~d lifetime for every star, so the activity-driven
false-alarm channel is sampled at one point of the distribution, not bounded.
\emph{(iii)}~Its systematics are the residuals left after the planned mission
corrections (drift correction and microscan assumed applied; the end-of-life
charge-transfer-inefficiency correction not yet quantified), treated here as
independent between cameras, whereas the dominant common-mode terms such as
pointing jitter are correlated within a camera group \citep{Jannsen2024}, so
the real multicamera averaging is less favourable.
\emph{(iv)}~Pixel-level effects and blends are out of scope, so blended
astrophysical false positives are not probed.

The forecast's model limits are:
\emph{(1)}~the occurrence rates dominate the field-population interval (a factor
of two); \emph{(2)}~the noise interpolation, anchored at $P$-band magnitudes
13.5--15.0, is extrapolated brightward for 15 of the 35 HZ hosts, which carry
about 80\% of the integrated probability, most of them on the completeness
plateau; \emph{(3)}~activity is fixed at the cells' median level;
\emph{(4)}~the 18- and 24-camera zones (24\% of the field) use a
$1/\sqrt{N_{\rm cam}}$ scaling valid only when photon-limited;
\emph{(5)}~the HZ forecast extends the completeness curve beyond its measured
$P\leq100$~d domain for about a third of the hosts (Sect.~\ref{sec:forecast});
\emph{(6)}~the limb-darkening window factor is held constant over impact
parameter ($-1.7$ points on the $1.0\,R_\oplus$ forecast); \emph{(7)}~binaries
are unfiltered and orbits circular; \emph{(8)}~the coherence test's
per-event timing degrades for transits longer than about 4~h, a PLATO regime
inherited from TESS and deferred to a future version; and \emph{(9)}~the
injections sample impact parameters uniformly over $0$--$0.6$ whereas the
forecast draws them over $0$--$0.9$, so about a third of the forecast
population lies outside the range on which the completeness was measured, in
the unfavourable direction. The quoted $41\pm1\%$ reflects the
completeness-curve statistical width alone; the model limits above point in
both directions, the camera scaling, the period extrapolation and the brightward
noise extrapolation making it an over-estimate, without their relative weights
being established here. The completeness and yield are scored
at the frozen pink-S/N floor of 7; at the adopted discovery floor of 7.5 they
are a few points lower.

\section{Conclusions}\label{sec:conclusions}

An end-to-end injection-recovery campaign on simulated PLATO photometry of
M dwarfs yields three measured results. First, a pipeline validated on TESS
recovers 50\% of injected planets at an expected ${\rm S/N}=7.7\pm0.15$, and
the long-period regime that contains the habitable zones is measurably harder
($8.6\pm0.3$), so an HZ forecast uses the completeness curve of the
corresponding period regime. Second, the false-alarm rate on planet-free light
curves is a strong function of the red-noise-aware acceptance floor, falling
from $79\pm2$\% to 0.6\% between floors of 6 and 8; completeness and
reliability are therefore only meaningful as a pair, and the measured pair
motivates an operating floor of ${\snrpink}\approx7.5$ for discovery
($6\pm1$\% false alarms at $54\pm2$\% completeness) or 8 for statistical
purity (0.6\%), floors that the systematics arm leaves unchanged.
Third, applied to the real LOPS2 M dwarfs, the known transiting TESS objects of
interest of the field would all be recovered, while a transiting
habitable-zone Earth would be
recovered as a candidate about two in five times ($\sim$40\%), placing these
planets at the detection frontier, where gains in pipeline completeness map
directly onto the expected number of candidates; both deciding quantities can
be measured, as done here, before launch.

\section*{Data availability}

The scored per-planet and per-twin tables of both arms, the generator scripts
with the PSLS configurations and random seeds that regenerate the 1\,720 light
curves and the residual templates of the systematics arm, and the scripts that
produce the figures are available at
\url{https://gitlab.com/yohanntschudi/plato-p4-completeness}, together with the
detection chain at the frozen version used for the campaign. PSLS is
distributed by its authors \citep{Samadi2019}. The LOPS2 field population was
drawn from the all-sky PLATO input catalogue \citep{Montalto2021} through
VizieR.

\section*{Acknowledgements}
The author thanks R. Samadi for clarifying the magnitude domain of the
instrument-systematics tables of PSLS.
This work used PSLS~1.9, \texttt{transitleastsquares}~1.32,
\texttt{celerite2}~0.3.2, \texttt{numpy}~2.3.5, \texttt{scipy}~1.16.3,
\texttt{pandas}~2.3.3, \texttt{astropy}~7.2.0, and \texttt{matplotlib}~3.10.8.
The all-sky PLATO input catalogue (asPIC~1.1) was queried through VizieR (CDS,
Strasbourg). The pipeline and analysis were developed by the author with
assistance from Claude (Anthropic); all code, results, and text were reviewed
and validated by the author.

\bibliographystyle{aasjournal}
\bibliography{references}

\end{document}